\documentclass[english,twocolumn,aps,prl, superscriptaddress,showpacs, amsmath]{revtex4-2}

\usepackage{color,xcolor}

\usepackage{natbib}
\usepackage[T1]{fontenc}
\usepackage{units}
\usepackage{amssymb,amsmath}
\usepackage{graphicx}
\usepackage{esint}
\usepackage{bm}
\usepackage{ulem}
\usepackage[colorlinks=true, citecolor=blue]{hyperref}
\setcitestyle{journalcolor=blue}

\begin{document}

\title{Proximitized insulators from disordered superconductors}

\author{Moshe Haim}
\address{Department of Physics, Jack and Pearl Resnick Institute and the Institute of Nanotechnology and Advanced Materials, Bar-Ilan University, Ramat-Gan 52900, Israel}
\address{Department of Condensed Matter Physics, Weizmann Institute of Science, Rehovot 76100, Israel}
\author {David Dentelski}
\address{Department of Physics, Jack and Pearl Resnick Institute and the Institute of Nanotechnology and Advanced Materials, Bar-Ilan University, Ramat-Gan 52900, Israel}
\author{Aviad Frydman}
\address{Department of Physics, Jack and Pearl Resnick Institute and the Institute of Nanotechnology and Advanced Materials, Bar-Ilan University, Ramat-Gan 52900, Israel}

\date{\today}

\begin{abstract}
We present an experimental study of bilayers of a disordered $Ag$ metal layer close to the metal-insulator transition and an Indium Oxide film which is on the insulating side of the superconductor-insulator-transition. Our results show that superconducting fluctuations within the indium-oxide film, that proximitize the underlying metal layer, induce \textit{insulating} rather than superconducting behavior. This is ascribed to suppression of density of states (due to the superconducting energy gap) for quasiparticles in the proximitized regions. Our results present a novel manifestation of the proximity effect phenomenon and provide important insight into the nature of the insulating phase of the disorder driven superconductor-insulator-transition.  
\end{abstract}

\maketitle

The interplay between superconductivity and disorder is a very active topic of investigation. It was recognized decades ago that s-wave superconductivity is remarkably robust against weak disorder \cite{anderson1958absence}. The situation is different for strong disorder.  Experiments show that superconductivity in 2D films can be destroyed by strong enough disorder as well as other non-thermal tuning parameters, g, such as magnetic field, thickness, chemical composition, gate voltage and pressure \cite{strongin1970kammerer, dynes1986breakdown, haviland1989onset, valles1992electron, frydman2002universal, aubin2006magnetic, stewart2007superconducting, sacepe2008disorder, marrache2008thickness, hollen2011cooper, postolova2017reentrant, baturina2011nanopattern,poran2017quantum,shahar1992superconductivity,sacepe2011localization,poran2011disorder,roy2018quantum,paalanen1992low,yazdani1995superconducting,gantmakher1998destruction,sambandamurthy2004superconductivity,sambandamurthy2005experimental,steiner2005possible,baturina2005quantum,baturina2007quantum,crane2007fluctuations,vinokur2008superinsulator,ganguly2017magnetic,mondal2011phase,parendo2005electrostatic,caviglia2008electric,cohen2021setup}.  Once superconductivity is destroyed, the system undergoes a  transition to an insulating state (for reviews see \cite{goldman1998superconductor,lin2015superconductivity}). This superconductor-insulator-transition (SIT) is a paradigmatic example for a quantum phase transition that can occur in systems driven by a non-thermal tuning parameter \cite{sachdev2007quantum}. 

One of the ongoing deliberations in the field of the SIT is the nature of the insulating phase, $I_S$. It has been shown both theoretically \cite{ghosal1998role,ghosal2001inhomogeneous,bouadim2011single, dentelski2018tunneling}  and experimentally \cite{sherman2012measurement} that in some nominally s-wave BCS superconductors, the presence of disorder can separate the temperature $T^*$ where pairing occurs accompanied by the development of an energy gap in the local density of states \cite{sacepe2011localization,kamlapure2013emergence}, and the actual $T_c$ where the superfluid density becomes finite. The pseudogap region between these temperatures grows with disorder as the SIT is approached and eventually, in the insulating regime, a finite superconducting gap, $\Delta$, exists even in the absence of superfluid density.  Indeed, similar energy gap \cite{sherman2012measurement} (as well as vortex motion \cite{poran2011disorder,kopnov2012little} and Nernst signals \cite{roy2018quantum}) were measured in both the superconductor and $I_S$ phases of disordered films. These results led to the realization that a BCS superconductor can undergo a quantum phase transition to an insulating phase of bosons rather than unpaired electrons. Hence, $I_S$, which is composed of uncorrelated superconducting fluctuations, shows a number of properties similar to those of a bulk superconductor despite being an electrical insulator. In this letter, we study a special aspect of the superconducting nature of $I_S$, i.e. the proximity effect to a normal metal.

The classic proximity effect describes the mutual influence of two "clean" layers, one is a superconductor, $SC$, and one is a normal metal, $N$, placed in a good electric contact, resulting in the induction of finite superconductivity into the $N$ and suppression of the superconducting order parameter in the $SC$ \cite{de1964boundary,deutscher1969pg}. Experiments have shown \cite{merchant2001crossover,frydman2001granular} that, when the superconductor is highly disordered, superconducting quantum fluctuations (even within the $I_S$) can induce superconductivity into a proximitized $N$. In this letter we report on a more exotic effect which occurs when the $N$ is highly disordered so that it is close to the metal-insulator-transition. Our main results are the following: 
\begin{enumerate} 
\item  Placing a highly disordered metal in proximity to an $I_S$ can induce $\it{insulating}$ behavior in the $N$.

\item This effect becomes more prominent as the $I_S$ is driven towards the SIT.

\item The effect is larger the more disordered is the $N$ layer.
\end{enumerate} 
We provide a simple model to explain these results based on enhanced electronic localization in the disordered metal due to proximitized superconducting fluctuations in the normal region. 

\begin{figure}[h]
\begin{center}
\includegraphics[width=0.65\textwidth]{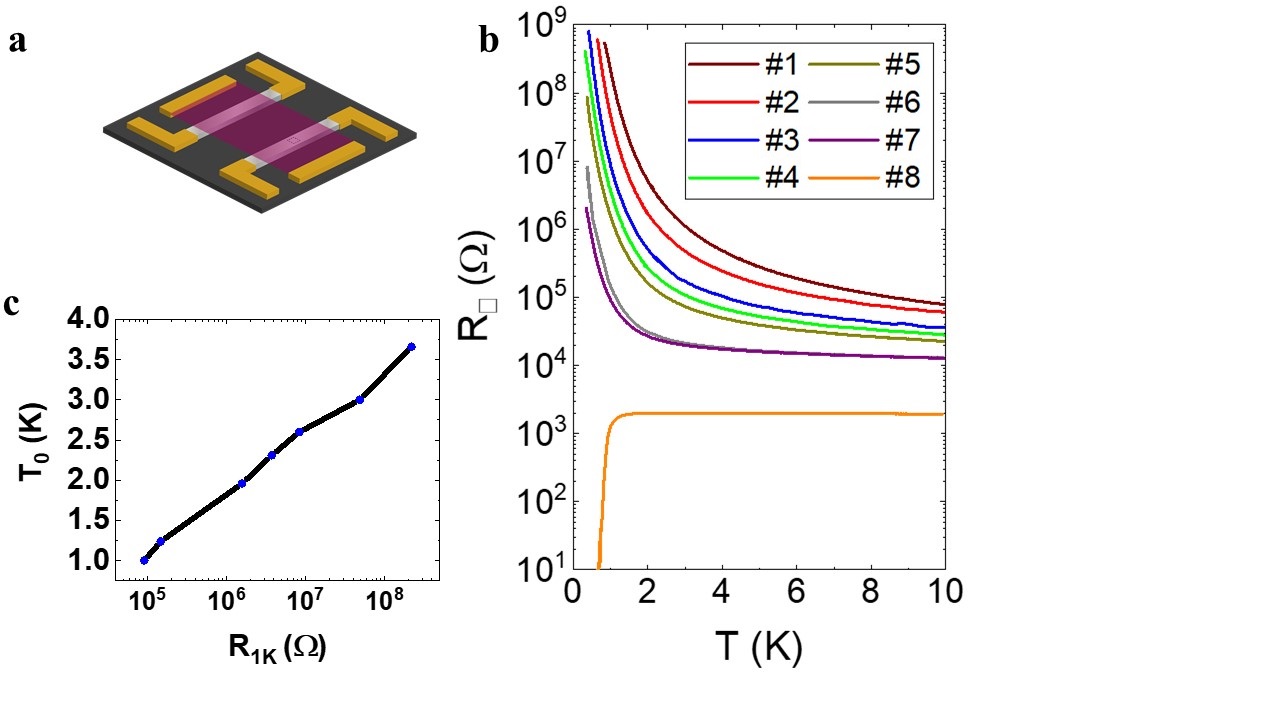}
\small{\caption{\textbf{a}: A sketch of the device containing six leads (gold), two silver strips (gray) and the $InO$ layer (purple). \textbf{b}: Resistance vs temperature of the $InO$ layer of sample $S1$, for different stages of annealing (as noted in the legend), measured between the silver strips. \textbf{c}: $T_0$ (extracted from the $R=R_{0} e^{\frac{T_0}{T}}$ dependency) vs the 1K $InO$ sheet resistance. The dashed line is a guide to the eye.
\label{INO}}}
\vspace{0cm}
\end{center}
\end{figure}

The samples for this study were prepared using the following scheme. Six $Au$ leads were deposited on an insulating $SiO$ substrate (gold pads in Fig \ref{INO}a). Then, two $10nm$ thick $Ag$ strips were deposited between two sets of leads, to be used as the  $N$ proximity layer (grey strips in Fig. \ref{INO}a). In order to increase the disorder of the silver, the samples were thinned in an $Ar$ plasma chamber in short pulses for different amounts of time. Here we present results for three highly disordered $Ag$ films, $S1$, $S2$ and $S3$, having decreasing $Ag$  room temperature sheet resistances of 250, 220 and 150 $\Omega_{\square}$, respectively.  Finally, a $30nm$ thick layer of amorphous Indium Oxide ($InO$) was e-beam deposited in a $10^{-4} mbar$ partial oxygen pressure resulting in a highly disordered, insulating film (purple layer in Fig \ref{INO}a). The resistance of the $InO$ film was sequentially  reduced via low temperature thermal annealing, thus driving the film through the insulator to superconductor transition \cite{ovadyahu1986some,roy2018quantum}. This setup takes advantage of the fact the $Ag$ layer is significantly more conductive than the $InO$ layer and allows to use the Ag layer as a voltage terminal for four-probe resistance measurements of the bare $InO$ film and two-probe measurements of the resistance of the $Ag/InO$ bilayer as the $InO$ is driven through the SIT.

$InO$ films, despite being morphologically uniform \cite{ovadyahu1986some,givan2012compositional}, have been shown to include emergent granularity in the form of superconducting puddles embedded in an insulating matrix \cite{kowal1994disorder,sherman2012measurement,kopnov2012little,roy2018quantum}. Hence, local superconductivity is present even in the insulating phase of the SIT.  Fig. \ref{INO}b shows the resistance vs temperature curves of the bare $InO$ of $S1$  for different annealing stages (sequentially reducing $R_{\square}$). The insulating curves are found to follow  $R=R_0e^{T_0/T}$ behavior. This is a typical feature of $I_S$ insulators which are characterized by emergent granularity \cite{shahar1992superconductivity,baturina2007localized,baturina2008hyperactivated,humbert2021overactivated}. Fig. \ref{INO}c presents $T_0$ versus $R$ of the $InO$ film, which is found to decreases as the sample approaches the SIT and extrapolates to zero close (but beyond) to it, in consistence with previously reported works \cite{shahar1992superconductivity}.

\begin{figure}[h]
\begin{center}
\includegraphics[width=0.65\textwidth]{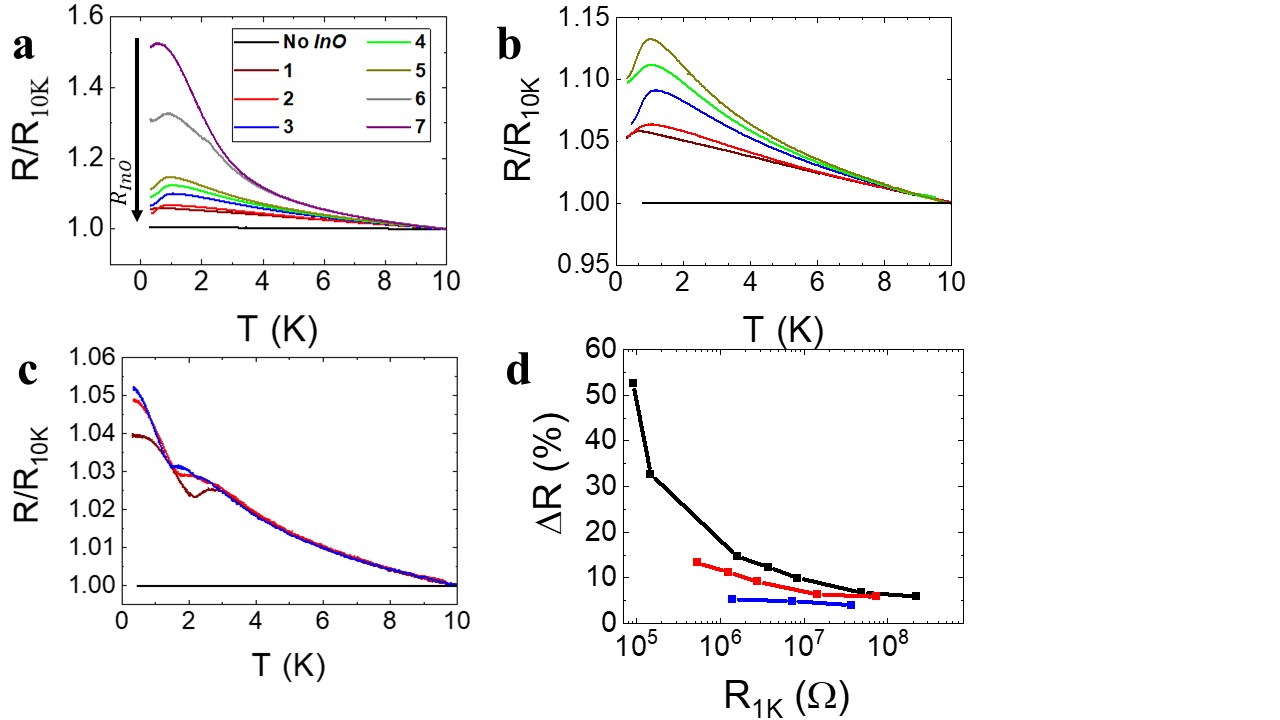}
\small{\caption{ Resistance vs temperature of the  three $Ag/InO$ bilayer of samples $S1$, $S2$ and $S3$ (panels \textbf{a}, \textbf{b} and \textbf{c}, respectively), for different stages of annealing. The colors signaling the annealing stage (see legend) apply for S2 and S3 as well, however, the sheet resistances of the InO film may differ between the samples. Black lines are plots for the bare $Ag$ films.  For clarity, the curves are normalized to the resistance at 10K. \textbf{d}: Resistance maximum  vs the $InO$ Resistance at 1K of samples S1 (black), S2 (red) and S3 (blue).    
\label{AG/INO}}}
\vspace{-0.5cm}
\end{center}
\end{figure}

The key result of this work is presented in Fig. \ref{AG/INO} which shows the resistance versus temperature curves for the three $Ag/InO$ bilayers. As long as the $InO$ is in the insulating phase, the resistance of such a bilayer is governed by the $Ag$ layer which has a much lower sheet resistance ($\approx 200 \Omega_{\square}$) than that of the $InO$ ($\approx 100 k\Omega_{\square}-100 M\Omega_{\square}$) for $T \leq 10K$. Thus, when measuring the bilayer we are, in fact, measuring the Ag layer almost strictly. We note that the annealing process may slightly affect the Ag resistance as well. 

We start with considering the results of sample  $S1$ (having the highest $Ag$ resistance) represented in Fig. \ref{AG/INO}a. Surprisingly, the addition of $InO$ causes the resistance to \textit{increase} with decreasing temperature below $\sim 10K$. This is very counter-intuitive since one naively expects that the $InO$ would add conductivity in parallel and, if anything, would reduce the total resistance. Instead, the $InO$ overlayer seems to be inducing insulating behavior in the underlying $Ag$ film.  This effect gets larger as the $InO$ is driven towards the SIT, eventually  exceeding a $50\%$ amplitude increase before reversing the trend at low temperatures where the resistance starts decreasing with lowering temperature.  

A similar effect, though with smaller magnitude,  was seen for samples $S2$ and $S3$, having decreasing disorder respectively (Fig. \ref{AG/INO}b and c).  Fig \ref{AG/INO}d shows the peak amplitude versus $R_{\square}$ for the three samples, indicating that the resistance increase depends on two parameters. It is larger the more disordered the  $Ag$ film is and also  the closer the $InO$ is to the  SIT. 

\begin{figure}[t]
\begin{center}
\vspace{-0cm}
\includegraphics[width=0.37\textwidth]{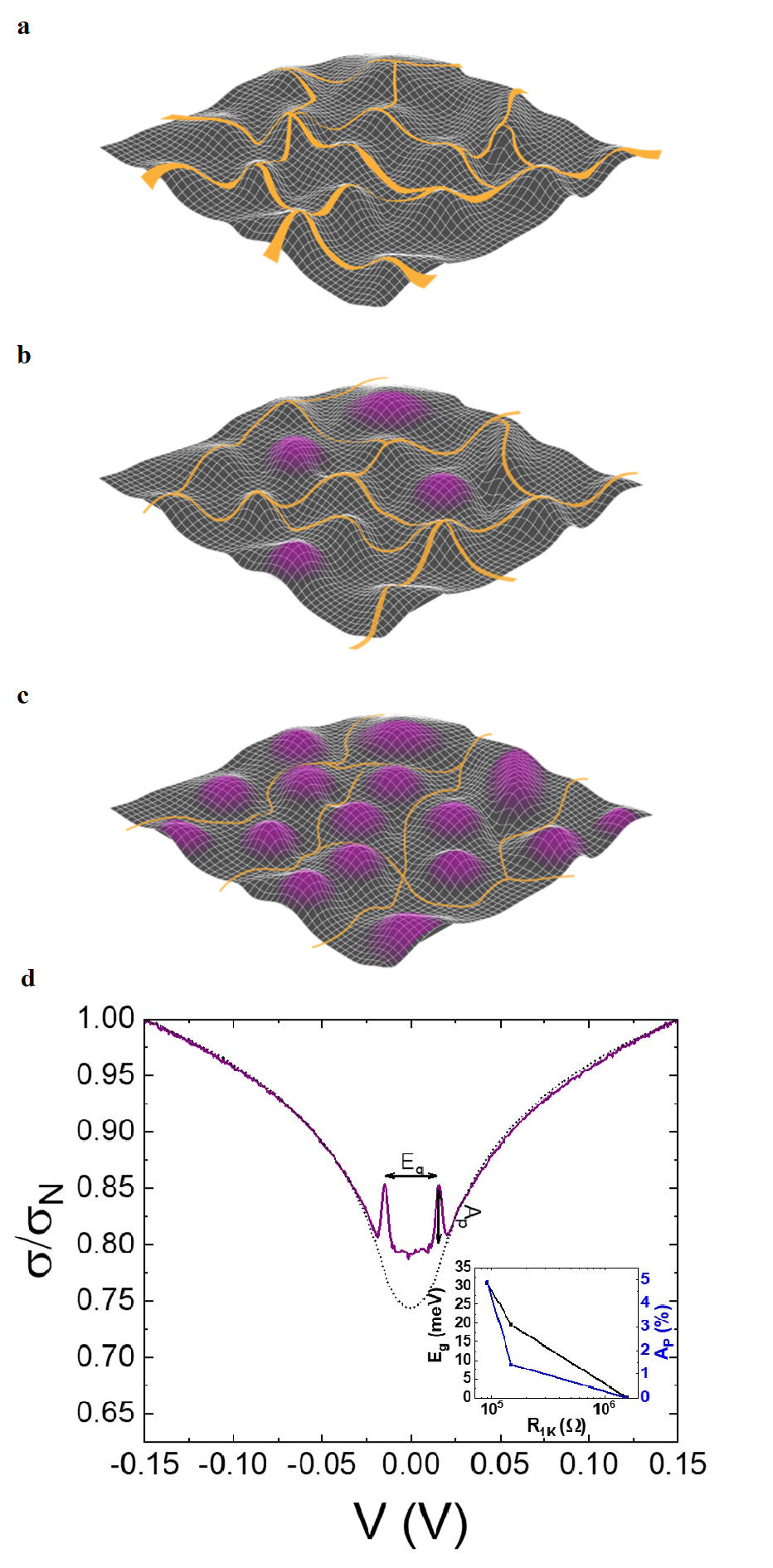}
\vspace{0cm}
\small{\caption{Illustration of the current paths (in orange) through the $Ag$ film.  \textbf{a}: The 2D conductivity map of the  $Ag$ film prior to the $InO$ deposition. The current is carried by the most conductive parts (peaks in the 2D map). \textbf{b}: Adding an $InO$ layer proximitizes different parts of $Ag$ film and induces superconductivity, represented in purple.  The current bypasses the SC regions due to suppression of the DOS in these locations. \textbf{c}: Annealing the $InO$ leads to more sections of the $Ag$ being proximitized thus further limiting the current paths and  forcing them to choose less conductive routes, thus further increasing the resistance. \textbf{d}: Differential conductivity vs bias voltage of sample $S1$ for the last insulating stage, normalized to the data at 150 meV. The inset shows the relative peak height and energy separation for different disorder degrees.
\label{model}}}
\vspace{-0.7cm}
\end{center}
\end{figure}

In order to understand how inducing superconducting fluctuations in a $N$ layer results in an increase of resistance we recall that our $Ag$ films are highly disordered, close to being insulating themselves. The conductivity is thus inhomogeneous due to strong spatial fluctuations of the underlying electronic potential. The current does not flow uniformly through the sample but rather through preferred high conductance trajectories as illustrated in the conductivity map of  Fig. \ref{model}a. 

Adding a disordered superconducting overlayer induces islands of $SC$ regions in the $Ag$ film (purple dots in Fig \ref{model}b) at temperatures below $T_c$. As the temperature is lowered, the density of superconducting regions increases. However, when global phase coherence is absent, the insertion of superconducting islands into a highly disorder metal can actually increase the resistivity. This is due to the formation of a local energy gap, $\Delta$, within each island, which suppresses the density of states for quasiparticles and thus limiting the current flow through these islands. Because the regions that are more prone to the proximity process are, naturally, those with higher conductivity, the current is restricted to one of two options:  flowing through the high resistance trajectories (Fig. \ref{model}b) or through the puddles of zero resistance, but at an energy 'cost' of $2\Delta$. Therefore, the sample can be viewed as a network of SIS junctions where the experimentally observed gap overlays the individual local ones which the current must tunnel through.   A similar process was suggested as the origin for the giant magnetoresistance peak observed in these materials at high fields and low temperatures \cite{dubi2006theory,dubi2007nature}.  

Lowering the disorder of the $SC$ and pushing it towards the SIT (e.g. by annealing the $InO$ layer) increases the density of superconducting puddles (Fig. \ref{model}c), thus further limiting the current carrying network and forcing the current to flow through higher resistance trajectories. This results in increasing the bilayer resistivity as the $InO$ film is pushed towards the SIT as indeed seen in the experiment (Fig. \ref{AG/INO}d).  In addition, reducing the $N$ disorder smooths the potential background thus suppressing the above process.

The temperature onset of this unique proximity effect is $\sim 10K$ which is significantly larger than the maximal $T_c$ measured in $InO$ films ($\sim 3.5K$ \cite{roy2018quantum}). However, STM measurements on a film of $InO$ with $T_C \approx 3K$ have detected a finite $\Delta$ up to temperatures of $\approx 6.5$ K \cite{sacepe2011localization}. In the insulator, $\Delta$ is predicted to grow further and increase as disorder increases \cite{bouadim2011single}. The real pairing critical temperature, $T^*$, of the $I_s$ phase of $InO$, is yet unknown, but the results presented here show signs for superconductivity up to $T \approx 10K$.

\begin{figure}[h]
\vspace{-0.2cm}
\begin{center}
\includegraphics[width=0.45\textwidth]{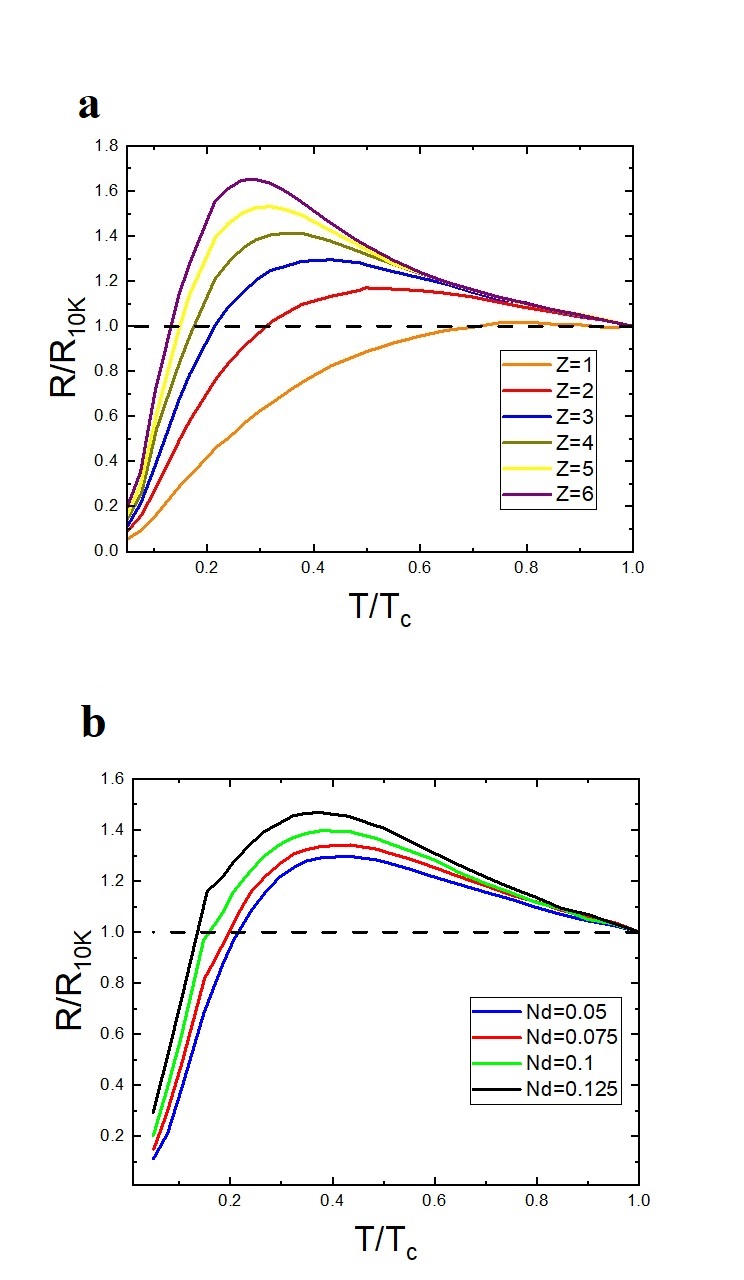}
\vspace{-0.4cm}
\small{\caption{Normalized sheet resistance $R_{\square}(t)$ of an $L = 50$ squared lattice normalized by its value at $T_{c}$, as a function of the reduced temperature $t = T/T_{c}$. \textbf{a}: Fixed disorder density $N_{d} = 0.05$ for different values of $Z$.  As the temperature decreases, there is an interplay between the gain of a current passing through a zero-resistance superconducting island and the cost, $Z$, to enter and exit the island. \textbf{b}:  Fixed $Z = 3$ for samples with different disorder density, $N_{d}$.
The dashed lines mark the sheet resistance at $T=T_c$.
\label{simulation}}}
\vspace{-0.5cm}
\end{center}
\end{figure}

The schematic representation of the current flow through the $Ag$ layer in Fig.~\ref{model}a-c is modeled here by considering an $L \times L$ squared lattice, where each site can be either a superconductor or a normal metal. We consider only nearest neighbors connections and assign a  resistance $r = 1$ for each bond between two metallic sites. The total sheet resistance, $R_{\square}$, is then calculated by the minimal resistance path required for the current to flow from one side of the bilayer to the other normalized by the size of the lattice. This is to say, that for $t=\frac {T}{T_c} \geq 1$, where all sites are metallic, $R_{\square}=1$.

With decreasing temperature, proximitized superconducting islands start to form in the $Ag$ layer. This is manifested by re-weighting all the bonds' resistances connecting a superconducting site to a metallic one by a factor $Z \geq 1$ while all bonds between two neighbouring superconducting sites (within a SC island) are assigned a resistance $r = 0$, such that $R_{\square} \rightarrow 0$ as the superconducting density, $n_{sc}$, becomes large. The different sites are chosen randomly to be metallic or superconductors, depending on the value of $n_{sc} (t) \in [0,1]$. 

We include the effect of disorder by introducing sites in the normal metal that prevent the formation of superconductivity. We assign a resistance $r(t) = r_{0}e^{T_0/T}$ between a site within a superconducting or a metal region to a disordered one. Here, we use $r_{0} = \exp(-1)$ and $T_{0} = 1.05 \  T_{c}$, such that the high-temperature resistivity of the different samples is $\approx 5\%$ higher than the clean one. The strength of the disorder is defined by the density of these sites, $N_d$. 

Fig. \ref{simulation}a shows the resistance as a function of the reduced temperature $t$ for a fixed value of $N_d$ and different values of $Z$, where we used the empirical approximate temperature dependence of the superconducting fraction:  $n_{sc}(t)=1-t^{0.4}$. For  all values of disorder, decreasing the temperature increases the density of superconducting islands and hence increasing $R_{\square}$.  This trend continues down to a disorder-dependent temperature, which marks the onset of global superconductivity, thus leading to the peak in the R-T curve as is found in the experiments. Note that the broad transition is also consistent with the experimental results. Fig. \ref{simulation}b shows the results for constant $Z=3$ and different degree of disorder,  $N_d$. It is seen that as the disorder increases, the resistivity peak is found to be higher as indeed observed in the experiments.  


The above picture is strongly supported by the differential conductance curve shown in Fig \ref{model}d which resembles an insulating stage of $S1$. The overall $dI/dV$  versus $V$ curve exhibits a suppression of conductance as the bias voltage is lowered, due to the Altshuler-Aronov (AA) mechanism of  electron-electron interactions \cite{altshuler1985electron} in the disordered $Ag$ film (dashed curve). At low bias, the curve exhibits a superimposed structure which includes two symmetrical maxima which resemble the coherence peaks of a superconducting gap structure. The peak amplitude and energy scale grow as the sample is further annealed and pushed towards the SIT, however they are only observed in the most disordered $Ag$ film ($S1$), and when the overlayer $InO$ film is close to the SIT, as seen in the inset. The energy scales extracted from these features are  19.6 and 30.8 meV for the two last insulating $InO$ stages.  Interestingly, these are integer multiples (14 and 22 respectively) of $1.4meV$, which is the value of $2\Delta$ for amorphous $InO$ \cite{sherman2012measurement}. This is consistent with the suggested model of current flowing through a series of $SIS$
junctions giving rise to a global gap-like structure, which is the sum of the individual local gaps on each of the SC island and is superimposed on the AA trend.

The results presented in this paper demonstrate a new type of proximity effect. We show that inducing superconductivity into an $N$ layer is not limited to 'clean' superconductors but can also be extended to an $I_S$ phase. Moreover, in the case of a highly disordered $N$, the proximity of a disordered metal to the $I_S$ can induce insulating-like behavior thus reducing its conductance. This effect becomes more prominent, the larger the N disorder. Such a proximitized bilayer can also offer a useful tool to study the $I_S$ deep into the insulating phase. Attempting to measure superconducting fluctuations in an insulating sample by transport is ineffective, since the exponentially increasing resistance screens local superconductivity. Tunneling measurement require a barrier that is much more resistive than the sample itself, limiting the measurement to samples that are close to the transition and at relatively high temperatures. In contrary, by coupling a film that is well within the insulating phase of the SIT to a normal metal, one can access transport and tunneling measurement of the coupled metal, regardless of how insulating the superconductor is. The interplay between the metal and the superconductor is quantified in our numerical model as a single parameter denoted as $Z$, which can be extracted directly from simple tunneling measurements and can be studied for different samples and materials.

We are grateful for technical help from I. Volotsenko, R. Cohen, Y. Stein, A. Fried, A. Roy and M. Laav and useful discussions with J. Ruhman, N. Trivedi and T. Baturina. This work was supported by the US-Israel Bi-national Science Foundation (BSF) grant No. 2020331.

\end{document}